\documentclass[twocolumn,showkeys,showpacs,preprintnumbers,prd,superscriptaddress,nofootinbib]{revtex4-1}
\usepackage{graphicx}
\usepackage{epsf}
\usepackage{bm}
\usepackage{amsmath}
\usepackage{amsfonts}
\usepackage{amssymb}
\usepackage{epstopdf}
\usepackage{natbib}
\usepackage{hyperref}
\usepackage{color}
\usepackage{verbatim}
\usepackage{multirow}
\usepackage{bm}
\usepackage{hyperref}
\usepackage{float}

\definecolor{darkblue}{rgb}{0.0, 0.0, 0.55}
\definecolor{darkred}{rgb}{0.55, 0.0, 0.0}

\usepackage{hyperref}
\hypersetup{
    colorlinks=true, 
    linkcolor=darkblue,
    citecolor=darkblue,
    urlcolor=darkblue}
    
\makeatletter\let\expandableinput\@@input\makeatother

\begin{document}

\title{New constraints on interacting dark energy from DESI DR2 BAO observations}

\author{Emanuelly Silva}
\email{emanuelly.santos@ufrgs.br}
\affiliation{Instituto de F\'{i}sica, Universidade Federal do Rio Grande do Sul, 91501-970 Porto Alegre RS, Brazil}

\author{Miguel A. Sabogal}
\email{miguel.sabogal@ufrgs.br}
\affiliation{Instituto de F\'{i}sica, Universidade Federal do Rio Grande do Sul, 91501-970 Porto Alegre RS, Brazil}

\author{Mateus Scherer}
\email{ mateusscherersouza@gmail.com}
\affiliation{Instituto de F\'{i}sica, Universidade Federal do Rio Grande do Sul, 91501-970 Porto Alegre RS, Brazil}

\author{Rafael C. Nunes}
\email{rafadcnunes@gmail.com}
\affiliation{Instituto de F\'{i}sica, Universidade Federal do Rio Grande do Sul, 91501-970 Porto Alegre RS, Brazil}
\affiliation{Divisão de Astrofísica, Instituto Nacional de Pesquisas Espaciais, Avenida dos Astronautas 1758, São José dos Campos, 12227-010, São Paulo, Brazil}

\author{Eleonora Di Valentino}
\email{e.divalentino@sheffield.ac.uk}
\affiliation{School of Mathematical and Phisical Sciences, University of Sheffield, Hounsfield Road, Sheffield S3 7RH, United Kingdom}

\author{Suresh Kumar}
\email{suresh.kumar@plaksha.edu.in}
\affiliation{Data Science Institute, Plaksha University, Mohali, Punjab-140306, India}

\begin{abstract}
In its second data release (DR2), the \textit{Dark Energy Spectroscopic Instrument} (DESI) publicly released measurements of Baryon Acoustic Oscillations (BAO) from over 13.1 million galaxies and 1.6 million quasars, covering the redshift range $0.295 \leq z \leq 2.330$. In this work, we investigate the impact of this new dataset on dark sector interaction models, which are motivated by non-gravitational interactions between dark energy (DE) and dark matter (DM), commonly referred to as interacting dark energy models (IDE). We focus on two frameworks: the traditional IDE model and the recently proposed sign-switching Interacting model (S-IDE), aiming to derive new and robust constraints on both scenarios. After carefully selecting the sample for the joint analysis, ensuring compatibility among the data without significant tension, our main results indicate that both models can alleviate the $H_0$ tension, reducing it to moderate tension approximately $2.7\sigma$. The IDE model shows compatibility with the latest $S_8$ constraints from cosmic shear surveys, while the S-IDE model predicts lower values of $S_8$, which align with alternative perspectives on the $S_8$ tension. For the traditional IDE model, we derive new bounds for the coupling parameter, marking the strongest constraints to date through geometric measurements. This highlights the crucial role that supernova samples can play in refining these constraints. For the S-IDE model, we find mild evidence (over $2\sigma$) for a non-zero coupling, once the PantheonPlus dataset calibrated with Cepheid-based magnitude measurements is included in the analysis.
\end{abstract}

\maketitle

\section{Introduction}

The $\Lambda$CDM model, long considered the standard of modern cosmology, is facing increasingly significant challenges as ever more precise observational data become available. Well-known tensions, such as those associated with $H_0$ and $S_8$, highlight significant discrepancies between different observational datasets (see~\cite{Perivolaropoulos:2021jda,Abdalla:2022yfr,DiValentino:2022fjm,Peebles:2024txt} for a review). For $H_0$, a divergence exceeding $5\sigma$~\cite{Verde_2019,Di_Valentino_2021,DiValentino:2021izs,Schoneberg:2021qvd,Shah:2021onj,Kamionkowski:2022pkx,Giare:2023xoc,Hu:2023jqc,Verde:2023lmm,DiValentino:2024yew,Perivolaropoulos:2024yxv} arises between early-universe measurements, based on the cosmic microwave background (CMB)~\cite{Planck:2018vyg}, and local determinations of the universe's expansion rate~\cite{Riess:2021jrx}. 
Regarding $S_8$, a key parameter that quantifies the amplitude of matter density fluctuations, a significant debate has emerged regarding a possible tension in its value when inferred from large-scale structure formation measurements and Planck-CMB data. Previous analyses of cosmic shear suggested a tension exceeding \( 3\sigma \)~\cite{KiDS:2020suj,DES:2017qwj}, but recent updates from the KiDS survey indicate excellent agreement with the $\Lambda$CDM model~\cite{Wright:2025xka,stölzner2025kids}. 
Measurements of $S_8$ from CMB data using ACT are also consistent with Planck-CMB predictions, reinforcing the standard cosmological model~\cite{louis2025atacama}. Other observational probes show agreement with $\Lambda$CDM at the $S_8$ level~\cite{Sailer:2024coh,Chen:2024vvk,Anbajagane:2025hlf,Garcia-Garcia:2024gzy,Sugiyama:2023fzm}. On the other hand, full-shape analyses of galaxy clustering data reveal a tension at the level of at least \( 4.5\sigma \)~\cite{Ivanov:2024xgb,Chen:2024vuf}, highlighting a potential discrepancy between early- and late-time universe probes. 
Furthermore, inferences based on Redshift-Space Distortions (RSD) could also exhibit significant tensions~\cite{Nunes_Vagnozzi_2021,Kazantzidis_2018}, adding another layer of complexity to the discussion. Other observational probes may also lead to a tension in $S_8$~\cite{Karim:2024luk,Dalal:2023olq}. The persistence of these discrepancies across different cosmological probes raises important questions about the possible need for extensions to $\Lambda$CDM, such as modifications to matter power spectrum modeling, alternative DE models, or interactions in the dark sector.

In addition to these tensions, several other anomalies and inconsistencies have been reported in the literature~\cite{Escudero:2024uea, Stahl:2025mdu, Escamilla:2025imi, Rogers:2023upm, Yin:2024hba, Wu:2024faw, deCruzPerez:2024shj, Bull:2015stt,Mukherjee:2025fkf,Pedrotti:2024kpn}. While they may have systematic origins, the persistence of these issues across different probes and cosmological studies suggests that multiple independent measurement teams would have to be committing multiple, unrelated errors~\cite{Riess:2019qba,DiValentino:2020vnx,DiValentino:2022fjm}. The possibility of this occurring is low, prompting us to seriously consider the need to explore new cosmological scenarios beyond the standard model.

In this context, various alternative models have emerged in the literature, aiming to restore observational concordance among cosmological parameters~\cite{DiValentino:2017oaw, FrancoAbellan:2020xnr, Banik:2024vbz, Adi:2020qqf, Feng:2019jqa, Giare:2024akf, Vagnozzi:2023nrq, Akarsu:2023mfb, Gomez-Valent:2024ejh, Toda:2024ncp, Yadav:2024duq, Gomez-Valent:2024tdb, Akarsu:2024eoo, Vagnozzi:2019ezj}. Among these, interacting dark energy (IDE) models stand out due to their simplicity and ability to reproduce the observational successes of $\Lambda$CDM while alleviating or resolving some of its tensions~\cite{Kumar:2016zpg, Murgia:2016ccp, Kumar:2017dnp, DiValentino:2017iww, Kumar:2021eev, Pan:2023mie, Benisty:2024lmj, Yang:2020uga, Forconi:2023hsj, Pourtsidou:2016ico, DiValentino:2020vnx, DiValentino:2020leo, Nunes_2021, Yang:2018uae, vonMarttens:2019ixw, Lucca:2020zjb, Zhai_2023, Bernui:2023byc, Hoerning:2023hks, Giare:2024ytc, Escamilla:2023shf, vanderWesthuizen:2023hcl, Silva:2024ift, DiValentino:2019ffd, Li:2024qso, Pooya:2024wsq, Halder:2024uao, Castello:2023zjr, Yao:2023jau, Santana:2023gkx, Mishra:2023ueo, Nunes:2016dlj, Teixeira:2024qmw, Giare:2024smz, Zhai:2025hfi, Li:2025owk, Li:2024qso, Feng:2025mlo, Lima:2024wmy, Yao:2024kex, Li:2024qso, Escamilla:2023shf,SantanaJunior:2024cug}. These models propose a direct, non-gravitational interaction between dark matter (DM) and dark energy (DE), regulated by a coupling parameter $\xi$. However, a central challenge of these models—and many others—is that the mechanism designed to alleviate one tension often exacerbates another~\cite{Sabogal:2024yha}. Recent studies indicate that IDE models can help mitigate the $H_0$ tension when the coupling parameter $\xi$ is negative (albeit worsening the $S_8$ tension), and reduce the $S_8$ tension when $\xi$ is positive (at the cost of exacerbating the $H_0$ discrepancy).

To address this issue,~\cite{Sabogal:2025mkp} recently proposed a variation of IDE models involving sign-switching interactions. Unlike traditional IDE scenarios, where the interaction term maintains a constant sign, this approach allows the direction of energy transfer between DM and DE to reverse over cosmic time. Notably, this new behavior is encoded in the coupling parameter itself, without introducing any additional degrees of freedom beyond those present in standard IDE frameworks, thus avoiding unnecessary parameter degeneracies. This phenomenon of signal switching has been explored in the context of effective fluid descriptions~\cite{Sabogal:2025mkp}, but similar behavior may also arise in the framework of type II minimally modified gravity theories~\cite{DeFelice:2020eju,DeFelice:2020cpt}, to which our model can be theoretically mapped by construction. Further investigation of such points will be presented in future work.

In this dynamic landscape, where various models are proposed to address cosmological tensions, it is essential to continuously update parameter constraints as new observational data become available. In this context, the Dark Energy Spectroscopic Instrument (DESI) plays a crucial role~\cite{DESI:2025fxa, DESI:2023ytc}, having recently released baryon acoustic oscillation (BAO) data based on millions of galaxies and quasars. DESI is a state-of-the-art spectroscopic survey designed to map the large-scale structure (LSS) of the universe in three dimensions, covering an extensive region of the sky and a broad range of redshifts. With its second data release (DR2)~\cite{DESI:2025fxa}, DESI has established itself as the largest spectroscopic dataset of its kind to date. These data enable the construction of 3D maps of the distribution of galaxies and quasars, the measurement of the universe's expansion~\cite{DESI:2025zgx, DESI:2024mwx}, the determination of the dark energy equation of state (EoS) parameter~\cite{DESI:2025kuo, DESI:2024aqx, Scherer:2025esj}, among other applications~\cite{DESI:2024hhd, Ishak:2024jhs, DESI:2025ejh, DESI:2024kob,Shah:2025ayl,Pang:2025lvh,Paliathanasis:2025dcr,Hur:2025lqc,DESI:2025hce,Anchordoqui:2025fgz,Pan:2025psn,Jiang:2024viw,Jiang:2024xnu,Chakraborty:2025syu,Pan:2025qwy,You:2025uon,Khoury:2025txd}.

The objective of this work is to use the DR2 survey data to derive new constraints on the dark sector, exploring both the traditional IDE model and the IDE sign-switching model, which introduces a more complex interaction dynamics between DM and DE. This \textit{paper} is structured as follows: Sec.~\ref{model} presents the two IDE models considered in this study. Sec.~\ref{data} describes the methodology and datasets employed in our analysis. The results are presented in Sec.~\ref{results}, and our final conclusions are summarized in Sec.~\ref{conclusions}. 

\section{Dark Sector Interaction Models}
\label{model}

In this section, we provide a brief review of interaction models between DE and DM, focusing on two main approaches. The first is the traditional model, where the interaction remains constant over cosmic time and the coupling parameter is constrained by prior assumptions. The second is the sign-switching model, which allows the direction of energy transfer between the dark components to reverse, enabling a broader investigation of the coupling parameter without restrictive priors.

\subsection{Traditional Approach}
\label{modelI}

In the IDE framework for describing interactions within the dark sector, a parameterization is introduced into the conservation laws, ensuring that while the total energy-momentum tensor remains conserved, the individual contributions of DE and DM do not satisfy independent conservation equations~\cite{Wang:2024vmw}. This condition is expressed as:

\begin{equation}
    \sum_{j} \nabla_\mu T_{j}^{\mu\nu} = 0, \quad \text{with} \quad \nabla_\mu T_{j}^{\mu\nu} = \frac{Q_j u_{\rm DM}^\nu}{a}.
    \label{eq:conservation}
\end{equation}

\noindent Here, the index $j$ represents DE or DM, $a$ denotes the scale factor, $u_{\rm DM}^\nu$ represents the four-velocity of DM, and $Q$ denotes the interaction rate between DE and DM.  
Starting from Eq.~\eqref{eq:conservation} and assuming a flat Friedmann–Lemaître–Robertson–Walker (FLRW) metric, we obtain the evolution equations for the energy densities of DM ($\rho_{\rm DM}$) and DE ($\rho_{\rm DE}$) in IDE models, given by:

\begin{eqnarray}
\frac{d\rho_{\rm DM}}{d\tau} + 3 \mathcal{H} \rho_{\rm DM} &=& Q, \\
\frac{d\rho_{\rm DE}}{d\tau} + 3 \mathcal{H} (1 + w_{\rm DE}) \rho_{\rm DE} &=& -Q.
\end{eqnarray}

\noindent In these expressions, $\mathcal{H}$ represents the conformal Hubble parameter, $w_{\rm DE}$ denotes the equation-of-state parameter for dark energy, and $Q$ corresponds to the same interaction term introduced in Eq.~\eqref{eq:conservation}. The choice of $Q$ is purely phenomenological, given our limited knowledge of the dark sector.\footnote{Several distinct parameterization proposals exist in the literature; see, for example,~\cite{Gavela:2010tm,DiValentino:2017iww,Zhai_2023,Caprini_2016,Liu_2022,Sharov_2016,Cid_2019,Feng_2016,Aljaf_2021,Zhai:2025hfi}.}

In this work, we adopt the interaction term $Q = \xi \mathcal{H} \rho_{\mathrm{DE}}$, where $\xi$ is a dimensionless parameter that governs the strength of the coupling between DE and DM. Within this framework, energy-momentum is transferred from DM to DE for $\xi < 0$, while the reverse occurs for $\xi > 0$. Notably, the parameter $\xi$ remains constant across any redshift interval, i.e., $\xi(z) = \text{const}$. This interaction model has been extensively studied in the literature in recent years (see~\cite{Kumar:2016zpg, Murgia:2016ccp, Kumar:2017dnp, DiValentino:2017iww, Kumar:2021eev, Pan:2023mie, Benisty:2024lmj, Yang:2020uga, Forconi:2023hsj, Pourtsidou:2016ico, DiValentino:2020vnx, DiValentino:2020leo, Nunes_2021, Yang:2018uae, vonMarttens:2019ixw, Lucca:2020zjb, Zhai_2023, Bernui:2023byc, Hoerning:2023hks, Giare:2024ytc, Escamilla:2023shf, vanderWesthuizen:2023hcl, Silva:2024ift, DiValentino:2019ffd, Li:2024qso, Pooya:2024wsq, Halder:2024uao, Castello:2023zjr, Yao:2023jau, Mishra:2023ueo, Nunes:2016dlj, Sabogal:2024yha}, for instance).

\subsection{Sign-Switching Coupling in the Dark Sector}
\label{modelII}

In this new approach proposed by~\cite{Sabogal:2025mkp}, we retain the flat FLRW metric, and the conservation equations follow the same principles as in the traditional scenario. The key difference lies in the functional form of $Q$, which now exhibits a dynamic behavior, with the interaction factor $\xi$ redefined as a time-dependent function. Mathematically, this is expressed as $Q = \xi(z) \mathcal{H} \rho_{\mathrm{DE}}$, where  
\begin{equation}
    \xi(z) = \xi_{i} \operatorname{sgn} \left[ z_{\rm eq,dark} - z \right].
    \label{xi}
\end{equation}

\noindent As discussed in detail in~\cite{Sabogal:2025mkp}, in this new approach, $\xi$ changes sign at the equipartition redshift $z_{\rm eq}$, where the ratio between the total matter and DE densities is unity, i.e., $\Omega_{\rm M}/\Omega_{\rm DE}(z_{\rm eq}) = 1$. The expression for $z_{\rm eq,dark}$ is given by
\begin{equation}
    z_{\rm eq,dark} = \left[ \left( \frac{\Omega_{{\rm M},0}}{\Omega_{{\rm DE},0}} + \frac{\xi_{i}}{3w_{\rm eff}} \right) 
    \left( 1 + \frac{\xi_{i}}{3w_{\rm eff}} \right)^{-1} \right]^{\frac{1}{3w_{\rm eff}}} - 1,
\end{equation}

\noindent where the effective equation of state is defined as  \noindent $w_{\rm eff} = w_{\rm DE} + \frac{\xi_{i}}{3}$. The constraints on $z_{\rm eq}$ can be robustly obtained through the analysis of cosmological data.

\noindent This new form of the coupling parameter also affects the perturbation-level equations, modifying the Boltzmann equations for DM and DE, which, in the synchronous gauge, take the following form~\cite{Sabogal:2025mkp}:

\begin{subequations}
\begin{eqnarray}
\dot{\delta}_{\rm c} & = & - \left[ \theta_{\rm c} + \frac{\dot{h}}{2} \right] + \frac{\rho_{\rm x}}{\rho_{\rm c}} \, \Xi(z) \\
& & + \, \Gamma(z) \left[\frac{3 \mathcal{H}}{k^2}  (w_{\rm x} - 1) + \left(\delta_{\rm x} - \delta_{\rm c}\right) \right]\, , \nonumber \\
\dot{\theta}_{\rm c} & = & -\mathcal{H} \theta_{\rm c} \, , \\ 
\dot{\delta}_{\rm x} & = & -\left(1 + w_{\rm x}\right)\left[ \theta_{\rm x} + \frac{\dot{h}}{2} \right] - \Xi(z) \\
& & - \dfrac{3\mathcal{H}}{k^{2}} \left(1 - w_{\rm x}\right) \left[ k^{2} \delta_{\rm x} + 3 \mathcal{H} \theta_{\rm x} \right]\, , \nonumber \\ 
\dot{\theta}_{\rm x} & = & 2 \mathcal{H} \theta_{\rm x} + \frac{k^2}{1 + w_{\rm x}} \delta_{\rm x} + \Gamma(z) \frac{\rho_{\rm c} \left( 2\theta_{\rm x} - \theta_{\rm c} \right)}{\rho_{\rm x}(1 + w_{\rm x})}\,, 
\label{EB_4}
\end{eqnarray}
\label{E_boltzmann}
\end{subequations}

\noindent where the source terms of interaction are given by
\begin{subequations}
\begin{eqnarray}
\Gamma & = & \xi_{i} \mathcal{H} \frac{\rho_{\rm x}}{\rho_{\rm c}} \operatorname{sgn}\left[z_{\rm eq} - z\right]\,, \\
\Xi & = & \xi_{i} \left[ \frac{3 \mathcal{H}^{2}}{k^2}(1 - w_{\rm x}) + \frac{k v_T}{3} + \frac{\dot{h}}{6} \right] \operatorname{sgn}\left[z_{\rm eq} - z\right]\,.
\end{eqnarray}
\label{interaction_terms}
\end{subequations}

\noindent In the absence of any interaction within the dark sector, the standard $\Lambda$CDM model is recovered. It is also important to note that the above equations reduce to the traditional cases when $\xi(z) = \xi_{i} = \text{const}$, with the imposition of suitable prior limits, as is well known. Throughout this text, we refer to this framework as \texttt{S-IDE}.

\section{Datasets and methodology}
\label{data}
To test the theoretical structures explored in this work, we implemented our model in the \texttt{CLASS} Boltzmann solver~\cite{Blas:2011rf} and used the \texttt{MontePython} sampler~\cite{Brinckmann:2018cvx, Audren:2012wb} to perform Monte Carlo analyses via Markov Chains (MCMC). Convergence was ensured by the Gelman-Rubin criterion~\cite{Gelman_1992}, with $R - 1 \leq 10^{-2}$ in all runs.

The cosmological parameters sampled in this analysis include the six standard parameters of the $\Lambda$CDM model. These six parameters are: the physical baryon density $\omega_{\rm b} = \Omega_{\rm b} h^2$, the physical dark matter density $\omega_{\rm c} = \Omega_{\rm c} h^2$, the angular scale of the sound horizon at the surface of last scattering $100\theta_{\mathrm{s}}$, the amplitude of the primordial scalar power spectrum $A_{\mathrm{s}}$, the spectral index of the primordial power spectrum $n_{\mathrm{s}}$, and the reionization optical depth $\tau_{\text{reio}}$.
In addition to these six standard parameters, we also sample the coupling parameter $\xi_{i}$ within the context of IDE models. The prior for $\xi_{i}$ varies depending on the scenario being analyzed.\footnote{For the sign-switching scenario, the range was based on~\cite{Sabogal:2025mkp}, where we tested an expansion of up to one order of magnitude and found no significant impact on the results, confirming its suitability.} The priors for these parameters are listed in Table~\ref{tab:priors}.

\begin{table}[h]
    \centering\caption{The cosmological priors used in this work apply to both the traditional IDE model and its extension, the Sign-Switching Coupling model (S-IDE). Flat priors are assumed over the ranges shown.}
    \renewcommand{\arraystretch}{1.3}
    \begin{tabular}{l c c}
        \hline
        \textbf{Parameter} & \textbf{IDE} & \textbf{S-IDE} \\
        \hline \hline
        $\Omega_{\rm b} h^2$ & $[0.0, 1.0]$ & $[0.0, 1.0]$ \\
        $\Omega_{\rm c} h^2$ & $[0.0, 1.0]$ & $[0.0, 1.0]$ \\
        $100\theta_{\mathrm{s}}$ & $[0.5, 2.0]$ & $[0.5, 2.0]$ \\
        $\ln(10^{10} A_{\mathrm{s}})$ & $[1.0, 5.0]$ & $[1.0, 5.0]$ \\
        $n_{\mathrm{s}}$ & $[0.1, 2.0]$ & $[0.1, 2.0]$ \\
        $\tau_{\text{reio}}$ & $[0.004, 0.8]$ & $[0.004, 0.8]$ \\
        $\xi_{i}$ & $[-1, 0]$ & $[-1.5, 1.5]$\\
        \hline \hline
    \end{tabular}
        \label{tab:priors}
\end{table}

All analyses were processed with the \texttt{GetDist} package,\footnote{\url{https://github.com/cmbant/getdist}} allowing the extraction of numerical results, including 1D posteriors and 2D marginalized probability contours.

The datasets used in the analyses are described below.
\begin{itemize}
\item \textit{Cosmic Microwave Background} (\textbf{CMB}): We use CMB temperature and polarization anisotropy measurements from Planck's 2018 legacy release~\cite{Planck:2018vyg}, including their cross-spectra. Specifically, we adopt the high-$\ell$ \texttt{Plik} likelihood for TT ($30 \leq \ell \leq 2508$), TE, and EE ($30 \leq \ell \leq 1996$), along with the low-$\ell$ TT-only ($2 \leq \ell \leq 29$) and EE-only ($2 \leq \ell \leq 29$) \texttt{SimAll} likelihood~\cite{Planck:2019nip}. We also incorporate CMB lensing measurements reconstructed from the temperature 4-point correlation function~\cite{Planck:2018lbu}. This dataset is referred to as \texttt{CMB}.

\item \textit{Baryon Acoustic Oscillations} (\textbf{DESI-DR2}): We utilize BAO measurements from DESI’s second data release, which includes observations of galaxies and quasars~\cite{DESI:2025zgx}, as well as Lyman-$\alpha$ tracers~\cite{DESI:2025zpo}. These measurements, detailed in Table IV of Ref.~\cite{DESI:2025zgx}, cover both isotropic and anisotropic BAO constraints over $0.295 \leq z \leq 2.330$, divided into nine redshift bins. The BAO constraints are expressed in terms of the transverse comoving distance $D_{\mathrm{M}}/r_{\mathrm{d}}$, the Hubble horizon $D_{\mathrm{H}}/r_{\mathrm{d}}$, and the angle-averaged distance $D_{\mathrm{V}}/r_{\mathrm{d}}$, all normalized to the comoving sound horizon at the drag epoch, $r_{\mathrm{d}}$. We also incorporate the correlation structure of these measurements through the cross-correlation coefficients: $r_{V,M/H}$, capturing the correlation between $D_{\mathrm{V}}/r_{\mathrm{d}}$ and $D_{\mathrm{M}}/D_{\mathrm{H}}$, and $r_{M,H}$, which describes the correlation between $D_{\mathrm{M}}/r_{\mathrm{d}}$ and $D_{\mathrm{H}}/r_{\mathrm{d}}$. This dataset is referred to as \texttt{DESI-DR2}.

\item \textit{Type Ia Supernovae} (\textbf{SN Ia}): We use the following recent samples of SN Ia:
\begin{enumerate}
    \item[(i)] \textbf{PantheonPlus and PantheonPlus\&SH0ES}: The distance modulus measurements from SN Ia in the PantheonPlus sample~\cite{pantheonplus} include 1701 light curves from 1550 distinct SN Ia events, spanning a redshift range of $0.01$ to $2.26$. We designate this dataset as \texttt{PP}. Additionally, we consider a version of this sample that uses the latest SH0ES Cepheid host distance anchors~\cite{Riess:2021jrx} to calibrate the absolute magnitude of SN Ia, rather than centering a prior on the $H_{0}$ value from SH0ES. This approach allows for more robust results, and this version of the dataset is referred to as \texttt{PPS}. 

    \item[(ii)] \textbf{Union 3.0}: The Union 3.0 compilation, consisting of 2087 SN Ia within the range $0.001 < z < 2.260$, was presented in~\cite{Rubin:2023ovl}. Notably, 1363 of these SN Ia overlap with the PantheonPlus sample. This dataset features a distinct treatment of systematic errors and uncertainties, employing Bayesian hierarchical modeling. We refer to this dataset as \texttt{Union3}.

    \item[(iii)] \textbf{DESY5}: As part of their Year 5 data release, the Dark Energy Survey (DES) recently published results from a new, homogeneously selected sample of 1635 photometrically classified SN Ia with redshifts spanning $0.1 < z < 1.3$~\cite{DES:2024tys}. This sample is complemented by 194 low-redshift SN Ia (shared with the PantheonPlus sample) in the range $0.025 < z < 0.1$. We refer to this dataset as \texttt{DESY5}.
\end{enumerate}
\end{itemize}

\section{Results}
\label{results}

The analysis of the results will be conducted in two complementary stages. Initially, we will examine the observational constraints within the framework of the conventional IDE scenario, followed by a detailed investigation of the predictions of the extended S-IDE model. To complement the analysis, we will present graphical representations of the DESI-DR2 data, clearly contrasting the observed cosmic distance relations with theoretical predictions, as well as the behavior of the cosmic expansion rate in comparison with the best-fitting models under study.

To evaluate the agreement between models and their compatibility with each dataset analyzed, we perform a statistical comparison with the $\Lambda$CDM scenario using, in addition to basic $\chi^2$ measurements, the Akaike Information Criterion (AIC)~\cite{Akaike:1974vps}. The AIC is defined as:
\begin{equation}
\text{AIC} \equiv -2 \ln L_{\text{max}} + 2N,
\end{equation}
where $L_{\text{max}}$ is the maximum likelihood of the model, and $N$ is the number of free parameters. A lower AIC value indicates a better model, balancing both goodness of fit and model complexity. Specifically, the AIC incorporates a penalty term for the number of parameters, discouraging the inclusion of unnecessary parameters that could lead to overfitting.

\subsection{IDE}

We begin our analysis with the traditional model, compiling our results in Table~\ref{tab:IDE}. We consider five distinct data combinations. The results show that the baryon density $\Omega_{\rm b} h^2$ remains stable across all analyses, while the dark matter density $\Omega_{\rm c} h^2$ becomes significantly better constrained with the inclusion of auxiliary datasets. The interaction parameter $\xi$ shows a negative value for the CMB+DESI-DR2 analysis ($\xi = -0.132^{+0.088}_{-0.063}$), suggesting only moderate evidence ($\sim 1\sigma$) for the interaction model, and a lower limit of $\xi > -0.254$ at 95\% CL. 
In the following, we perform a joint analysis combining CMB, DESI-DR2, and PPS data, which significantly improves the constraints on our parameter space. For this case, we obtain $\xi = -0.116^{+0.060}_{-0.050}$ at 68\% CL, with an associated Hubble constant of $H_0 = 69.61 \pm 0.44$ km/s/Mpc. This joint analysis provides a robust $\sim 2\sigma$ indication of a dark sector coupling, while also mitigating the Hubble tension to $\sim 3\sigma$. Although this scenario does not fully resolve the $H_0$ tension, it offers substantial statistical alleviation, highlighting the potential impact of dark sector interactions on cosmological parameter estimates.

\begin{table*}[htpb!]
\centering\caption{68\% confidence level (CL) constraints for the parameters of the IDE model. In the last rows, we present the quantities $\Delta \chi^2_{\text{min}} \equiv \chi^2_{\text{min (IDE)}} - \chi^2_{\text{min ($\Lambda$CDM)}}$ and $\Delta \text{AIC} \equiv \text{AIC}_{\text{IDE}} - \text{AIC}_{\text{$\Lambda$CDM}}$, which compare the model fits. Negative values for both differences indicate a preference for the IDE model over the $\Lambda$CDM model, while positive values favor the $\Lambda$CDM model.}
\renewcommand{\arraystretch}{1.5}
\resizebox{\textwidth}{!}{
\begin{tabular}{l||cc||ccc} 
\hline
\textbf{Parameter} & \textbf{CMB+DESI-DR2} & \textbf{CMB+DESI-DR2+PPS} & \textbf{CMB+DESI-DR2+PP} & \textbf{CMB+DESI-DR2+DESY5} & \textbf{CMB+DESI-DR2+Union3} \\ 
\hline \hline
$10^{2} \Omega_{\rm b} h^2$ & $2.253\pm 0.012$ & $2.259\pm 0.012$ & $2.255\pm 0.013$ & $2.253\pm 0.012$ & $2.254\pm 0.013$  \\
$\Omega_{\rm c} h^2$ & $0.1028^{+0.0097}_{-0.0069}$ & $0.1045^{+0.0068}_{-0.0054}$ & $0.1130^{+0.0045}_{-0.0017}$ & $0.1158^{+0.0021}_{-0.00076}$ & $0.1121^{+0.0052}_{-0.0019}$  \\
$100\theta_{s}$ & $1.04210\pm 0.00027$ & $1.04214\pm 0.00028$ & $1.04211\pm 0.00027$ & $1.04212\pm 0.00026$ & $1.04212\pm 0.00027$  \\
$\ln(10^{10} A_{s})$ & $3.051\pm 0.014$ & $3.052\pm 0.015$ & $3.054^{+0.013}_{-0.015}$ & $3.053\pm 0.015$ & $3.054^{+0.014}_{-0.016}$  \\
$n_{s}$ & $0.9703\pm 0.0032$ & $0.9713\pm 0.0033$ & $0.9712\pm 0.0032$ & $0.9714\pm 0.0032$ & $0.9710\pm 0.0031$  \\
$\tau_{\mathrm{reio}}$ & $0.0591\pm 0.0070$ & $0.0597^{+0.0066}_{-0.0076}$ & $0.0608^{+0.0066}_{-0.0077}$ & $0.0604\pm 0.0074$ & $0.0603^{+0.0066}_{-0.0080}$  \\
$\xi$ & $-0.132^{+0.087}_{-0.064}$ & $-0.116^{+0.060}_{-0.050}$ & $\xi > -0.0538$ & $\xi > -0.0220$ & $\xi > -0.0630$  \\
$H_0 \, [\mathrm{km/s/Mpc}]$ & $69.61^{+0.54}_{-0.67}$ & $69.61\pm 0.44$ & $68.82^{+0.27}_{-0.37}$ & $68.56^{+0.22}_{-0.26}$ & $68.91^{+0.29}_{-0.41}$  \\
$\Omega_{\rm m}$ & $0.260^{+0.025}_{-0.019}$ & $0.264^{+0.017}_{-0.015}$ & $0.288^{+0.012}_{-0.0055}$ & $0.2956^{+0.0064}_{-0.0033}$ & $0.285^{+0.014}_{-0.0061}$  \\
$S_8$ & $0.860^{+0.024}_{-0.040}$ & $0.850^{+0.020}_{-0.028}$ & $0.8242^{+0.0095}_{-0.018}$  & $0.8165^{+0.0084}_{-0.011}$ & $0.827^{+0.011}_{-0.018}$\\
\hline 
$\Delta \chi^{2}_{\mathrm{min}}$ & $1.1$ & $-2.20$ & $1.38$ & $2.32$ & $0.5$ \\
$\Delta \mathrm{AIC}$ & $3.1$ & $-0.20$ & $3.38$ & $4.32$ & $2.5$ \\
\hline \hline 
\end{tabular}}
\label{tab:IDE}
\end{table*}

\begin{figure*}[tpb!]
    \centering
    \includegraphics[width=0.48\textwidth]{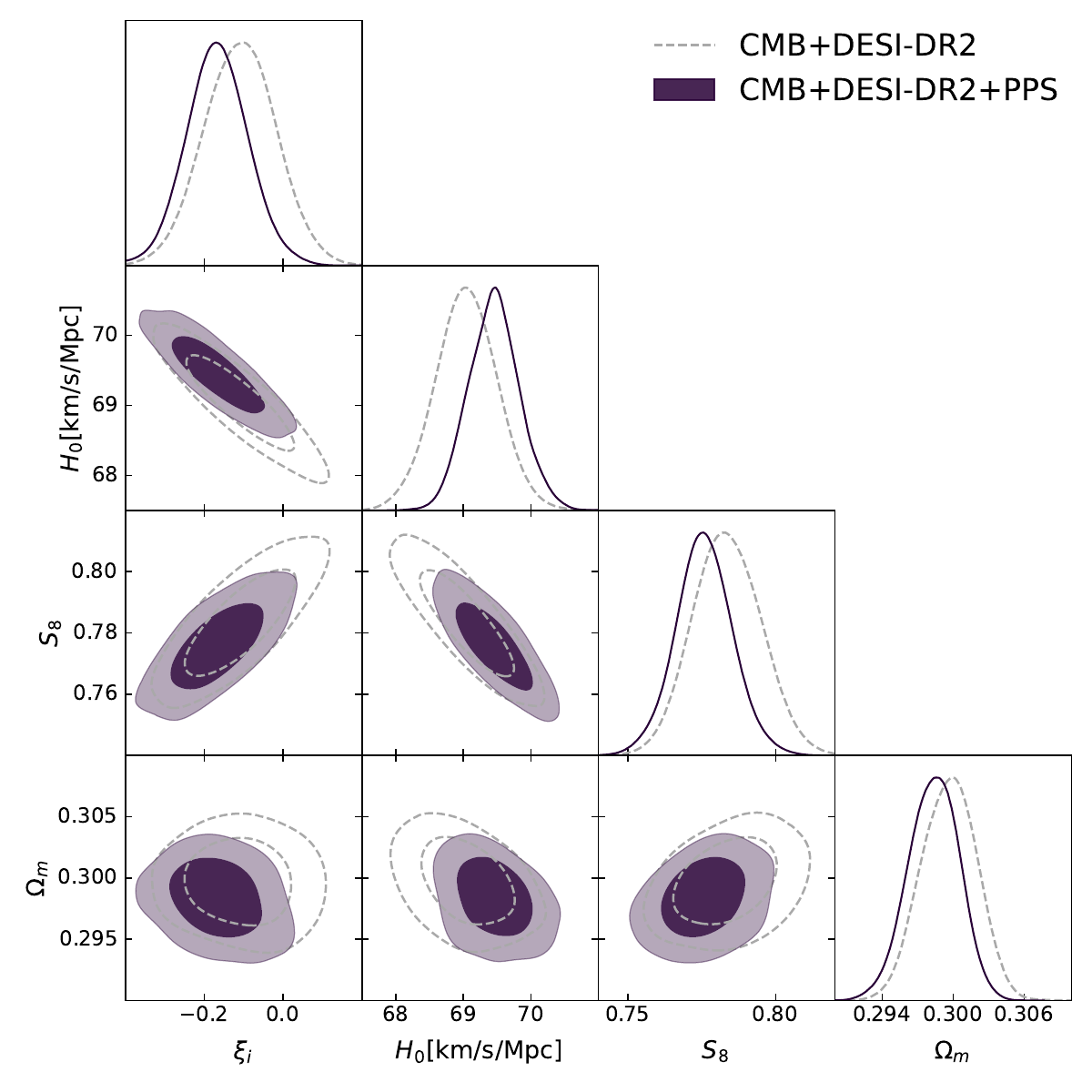}
    \includegraphics[width=0.48\textwidth]{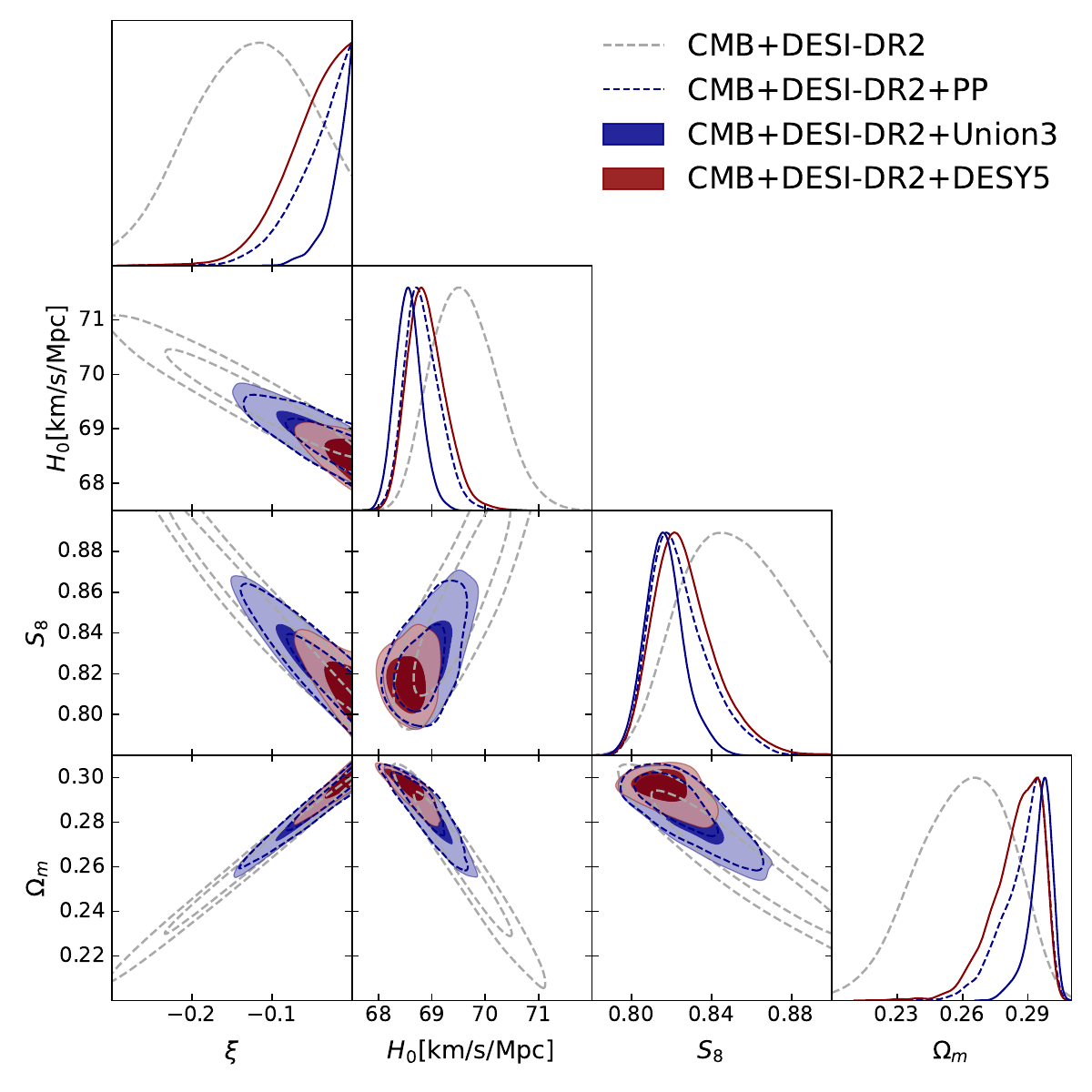} \hspace{0.02\textwidth}
    \caption{Marginalized posterior distributions and 68\% and 95\% CL contours for the parameters $\xi$, $H_{0}$, $S_{8}$, and $\Omega_{\rm m}$ in the S-IDE model (left panel) and IDE model (right panel). The results are shown for different combinations of datasets, as indicated in the legend.}
    \label{fig:IDE}
\end{figure*}

In the following, we proceed to evaluate the consistency of the model with respect to various SN Ia samples. Currently, three different SN Ia samples are publicly available in the literature: PP, Union3, and DESY5. The results for these analyses are summarized in Table~\ref{tab:IDE}, where different SN samples are combined with CMB+DESI-DR2. For all analyses performed, we find notably strong constraints on the model, imposing direct lower bounds on the coupling parameter: $\xi > -0.0538$, $\xi > -0.0220$, and $\xi > -0.0630$ when combined with PP, DESY5, and Union3, respectively. These analyses represent the strongest and most up-to-date constraints on $\xi$ to date, particularly for the joint CMB+DESI-DR2+DESY5 analysis.
The right panel of Figure~\ref{fig:IDE} summarizes our constraints on the parameter space at 65\% and 95\% CL for the cosmological parameters $\xi$, $H_{0}$, $S_{8}$, and $\Omega_{\rm m}$. It is observed that the precision of the measurements increases significantly with the inclusion of supernova data in the primary CMB+DESI-DR2 combination, with the addition of the DESY5 dataset providing the most precise constraints. Additionally, the values of $\Omega_{\rm m}$ remain consistently high and exhibit excellent agreement across the different data combinations (CMB+DESI-DR2 supplemented with PP, Union3, or DESY5).

\begin{figure*}[htpb!]
    \centering
    \includegraphics[width=0.96\columnwidth]{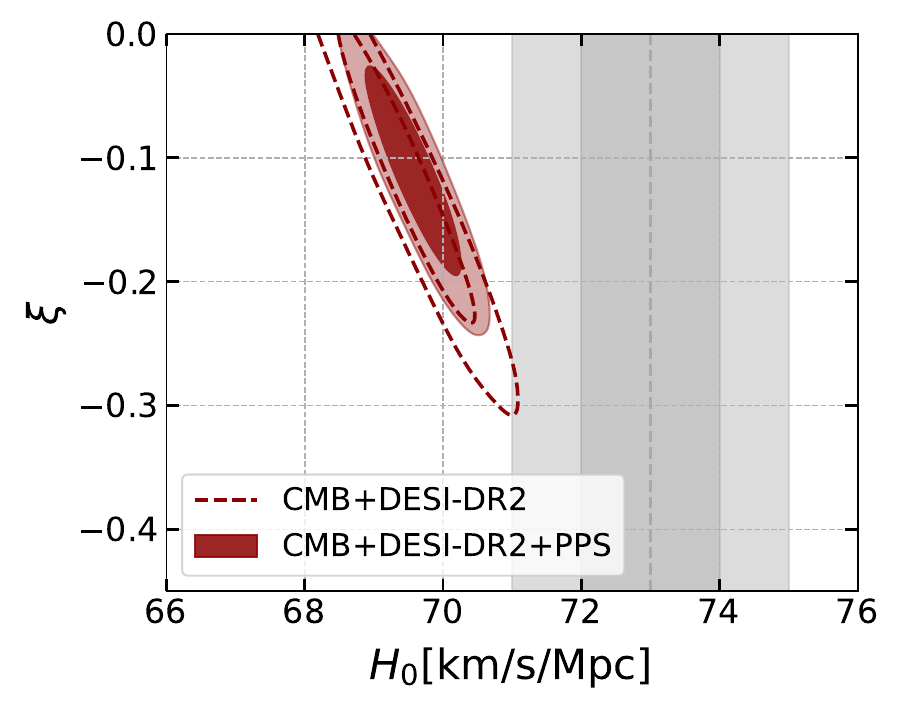} \,\,\,\
    \includegraphics[width=\columnwidth]{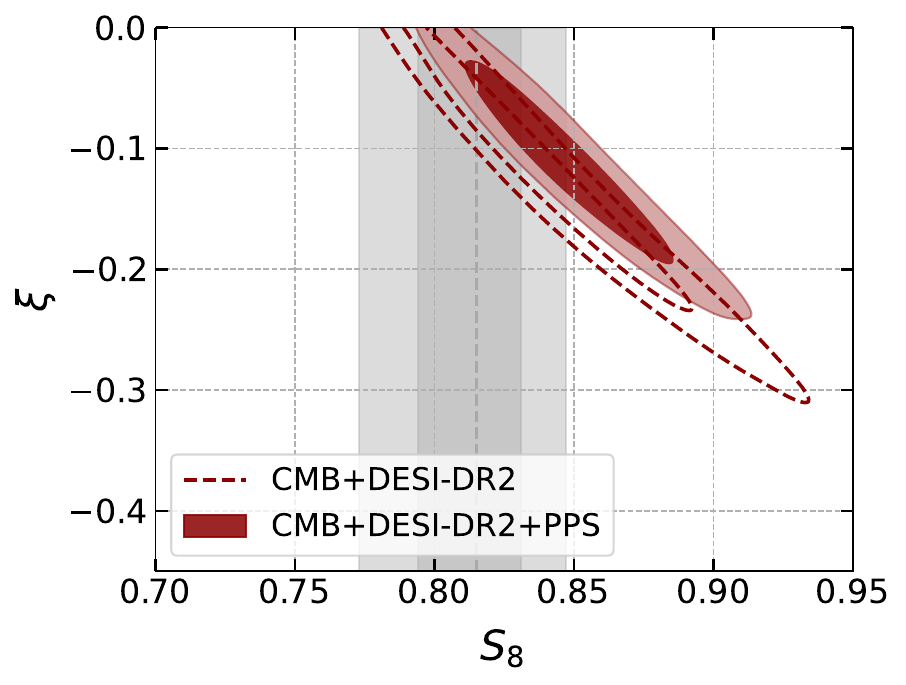} 
    \caption{Left panel: The 2D contours at 68\% and 95\% CL for the coupling parameter $\xi$ (traditional and usual IDE approach) and the Hubble parameter $H_0$ are derived from different combinations of CMB, DESI-DR2, and SN data, as indicated in the legend. The dark-gray band represents the $H_0$ value measured by the SH0ES collaboration ~\cite{Riess:2021jrx}: $H_0=73.04\pm1.04~{\rm km/s/Mpc}$. Right panel: Same as the left panel, but for the $\xi$-$S_8$ plane. The dark-gray band represents the $S_8$ value measured by the KiDS-Legacy survey \cite{Wright:2025xka}: $S_8\equiv\sigma_8\sqrt{\Omega_{\rm m}/0.3} = 0.815^{+0.016}_{-0.021}$.}
    \label{fig:S8ide}
\end{figure*}

Regarding the Hubble constant, all analyzed datasets yield values higher than those predicted by the $\Lambda$CDM model. Our constraints, derived from the CMB+DESI-DR2 and CMB+DESI-DR2+PPS combinations, reduce the tension with SH0ES measurements to $2.7\sigma$ and $3.1\sigma$, respectively. Although this reduction is less pronounced than that reported for DR1~\cite{Giare:2024smz}, there is an observed increase in precision of approximately $1\sigma$, representing a significant improvement in the constraints on this parameter within the model's framework. Figure~\ref{fig:S8ide} (left panel) shows the parametric space of $H_0$ versus $\xi$ as inferred from different combinations of CMB, DESI-DR2, and SN data, as indicated in the legend. The dark-gray band represents the value of $H_0$ measured by the SH0ES collaboration.

As for the $S_8$ parameter, the obtained values are consistent within the uncertainties with the latest measurements from the KiDS Legacy~\cite{Wright:2025xka}, as shown in the figure on the right panel of Fig.~\ref{fig:S8ide}. The inclusion of additional datasets—such as PP, DESY5, or Union3—leads to a slight reduction in the values, which nonetheless remain in statistical agreement with these observations. As shown in~\cite{Sabogal:2024yha}, analyses using RSD suggest systematically lower values for $S_8$, as predicted by RSD samples and full-shape galaxy clustering. Despite this discrepancy, the maximum level of tension (which occurs between the RSD measurements and the CMB+DESI-DR2 combination) does not exceed $< 2\sigma$, indicating low statistical significance.

It is important to emphasize that the incorporation of DESI DR2 data in this analysis provides significantly more precise constraints on the interaction parameter $\xi$ compared to previous results~\cite{Giare:2024smz}. Compared to DR1, the new measurements of the matter density parameter $\Omega_{\rm m}$ exhibit not only higher precision but also a systematic shift toward larger values, with uncertainties reduced by a factor of approximately two. From a physical perspective, since more negative values of $\xi$ imply a lower dark matter density, the observed increase in $\Omega_{\rm m}$ naturally favors a weaker interaction, driving $\xi$ closer to zero. This effect consistently explains the reduction in the statistical significance of the coupling compared to the results reported in~\cite{Giare:2024smz}.

\begin{figure*}[htpb!]
    \centering
    \includegraphics[width=0.96\columnwidth]{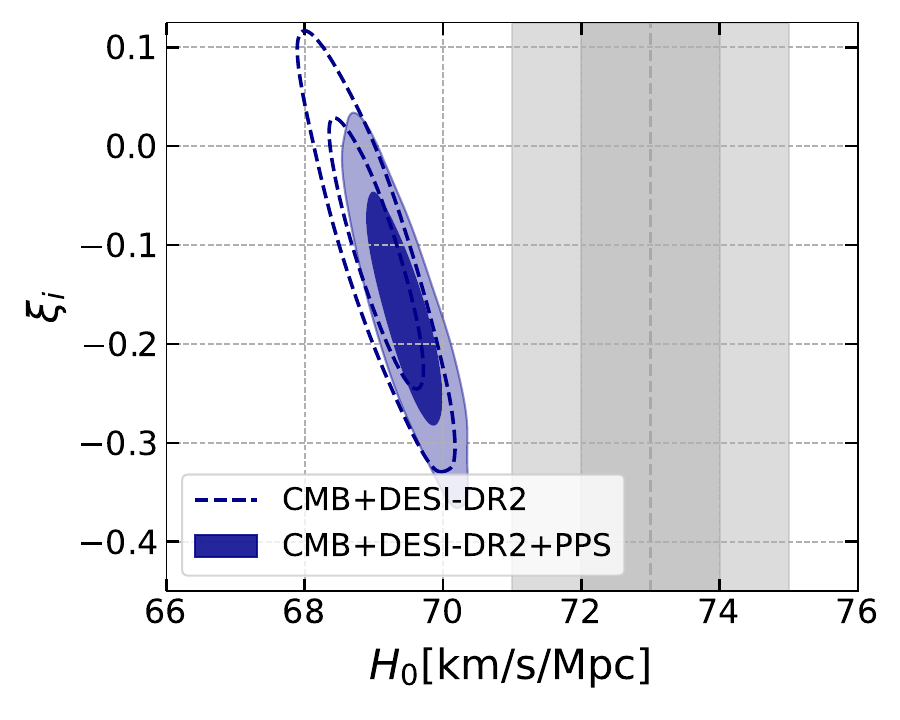} \,\,\,\
    \includegraphics[width=\columnwidth]{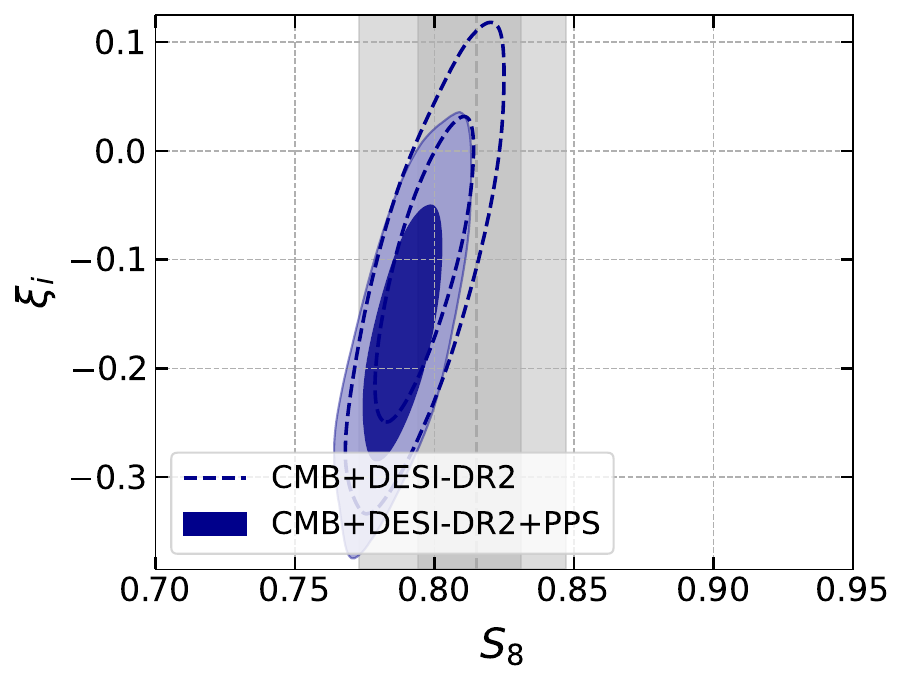} 
    \caption{Left panel: The 2D contours at 68\% and 95\% CL for the coupling parameter $\xi$ (S-IDE framework) and the Hubble parameter $H_0$ are derived from different combinations of CMB, DESI-DR2, and SN data, as indicated in the legend. The dark-gray band represents the $H_0$ value measured by the SH0ES collaboration~\cite{Riess:2021jrx}: $H_0=73.04\pm1.04~{\rm km/s/Mpc}$. Right panel: Same as the left panel, but for the $\xi$-$S_8$ plane. The dark-gray band represents the $S_8$ value measured by the KiDS-Legacy survey \cite{Wright:2025xka}: $S_8\equiv\sigma_8\sqrt{\Omega_{\rm m}/0.3} = 0.815^{+0.016}_{-0.021}$.}
    \label{fig:S8_Side}
\end{figure*}

\subsection{S-IDE}
Table~\ref{tab:Sign-Switching} summarizes our constraints on the cosmological parameters within the S-IDE framework. To ensure statistical robustness, we selected datasets that show no significant tension with the primary combination (CMB+DESI-DR2). The joint analysis CMB+PPS exhibits an almost negligible tension of only $0.10\sigma$ with this primary combination, whereas other datasets—such as CMB+PP, Union3, and DESY5—show significant discrepancies (reaching up to $4.64\sigma$ for DESY5), which could bias our analysis.\footnote{Tensions are quantified using the well-established estimator and methodology described in Section 3.A of~\cite{Sabogal:2025mkp}.} This methodological choice ensures a consistent exploration of the model parameters while minimizing contamination from systematic inconsistencies across datasets.\footnote{Notably, even within the $\Lambda$CDM framework, CMB+DESI-DR2 exhibits strong tensions with datasets such as PP, DESY5, and Union3, as previously reported in~\cite{DESI:2025zgx}.}

\begin{table}[htpb!]
\centering\caption{68\% confidence level (CL) constraints for the parameters of the S-IDE model. In the last rows, we present the quantities $\Delta \chi^2_{\text{min}} \equiv \chi^2_{\text{min (IDE)}} - \chi^2_{\text{min} (\Lambda\text{CDM})}$ and $\Delta \text{AIC} \equiv \text{AIC}_{\text{IDE}} - \text{AIC}_{\Lambda\text{CDM}}$, which compare the model fits. Negative values for both differences indicate a preference for the IDE model over the $\Lambda$CDM model, while positive values favor the $\Lambda$CDM model.}
\renewcommand{\arraystretch}{1.5}
\resizebox{\columnwidth}{!}{
\begin{tabular}{lcc} 
\hline
\textbf{Parameter} & \textbf{CMB+DESI-DR2} & \textbf{CMB+DESI-DR2+PPS} \\
\hline \hline
$10^{2} \Omega_{b} h^2$ & $2.253\pm 0.013$ & $2.257\pm 0.013$ \\
$\Omega_{c} h^2$ & $0.1197^{+0.0019}_{-0.0017}$ & $0.1207^{+0.0017}_{-0.0015}$\\
$100\theta_{s}$ & $1.04210\pm 0.00027$ & $1.04213\pm 0.00028$ \\
$\ln(10^{10} A_{s})$ & $3.051^{+0.014}_{-0.016}$ & $3.052^{+0.013}_{-0.015}$ \\
$n_{s}$ & $0.9707\pm 0.0033$ & $0.9706\pm 0.0034$ \\
$\tau_{\mathrm{reio}}$ & $0.0589^{+0.0067}_{-0.0079}$ & $0.0592\pm 0.0070$ \\
$\xi$ & $-0.109\pm 0.091$ & $-0.167\pm 0.080$\\
$H_0 \, [\mathrm{km/s/Mpc}]$ & $69.05\pm 0.46$ & $69.46\pm 0.38$\\
$\Omega_{\rm m}$ & $0.2996\pm 0.0023$ & $0.2983^{+0.0022}_{-0.0020}$ \\
$S_8$ & $0.796\pm 0.012$ & $0.7886\pm 0.0099$\\
$z_{\rm eq, dark}$ & $0.297^{+0.021}_{-0.028}$ & $0.287^{+0.017}_{-0.023}$ \\
\hline
$\Delta \chi^{2}_{\mathrm{min}}$ & $0.14$ & $-3.26$ \\
$\Delta \mathrm{AIC}$ & $2.14$ & $-1.26$\\
\hline \hline
\end{tabular}}
\label{tab:Sign-Switching}
\end{table}

In the joint CMB+DESI-DR2 analysis, we observe that the parameter $H_0$ assumes a lower value with improved precision compared to the results reported in~\cite{Sabogal:2025mkp}. The inclusion of the PPS dataset (CMB+DESI-DR2+PPS) further reduces the $H_0$ tension, bringing it close to the $3\sigma$ level. However, it is important to stress that this apparent alleviation is driven by the calibration of the PP dataset using SH0ES, which anchors the absolute magnitude of Type Ia supernovae with Cepheids. Without this calibration, the S-IDE model is generally disfavored relative to $\Lambda$CDM, as indicated by the $\Delta\chi^2$ and $\Delta\mathrm{AIC}$ values reported in Table~\ref{tab:Sign-Switching}. Figure~\ref{fig:S8_Side} further illustrates how the PPS dataset favors higher values of $H_0$.

Regarding the coupling parameter, the joint analyses CMB+DESI-DR2 and CMB+DESI-DR2+PPS yield $\xi = -0.109 \pm 0.091$ and $\xi = -0.167 \pm 0.080$, respectively, with the latter suggesting a moderate $\sim 2\sigma$ indication of a non-zero coupling. Nevertheless, this result should be interpreted with caution, as the inferred coupling strength is correlated with $H_0$, and its significance remains sensitive to the choice of dataset combinations.

Now, examining the $S_8$ parameter, we observe lower values compared to the traditional IDE scenario. However, this difference remains statistically insignificant when compared to the most recent cosmic shear results~\cite{Wright:2025xka}. For instance, the CMB+DESI-DR2+PPS combination exhibits only a $1\sigma$ tension with these measurements, as can be visually confirmed in the right panel of Figure~\ref{fig:S8_Side}. 
In this context, the S-IDE model offers significant advantages: in addition to maintaining compatibility with cosmic shear data, it substantially reduces tensions in $S_8$ between Planck measurements and RSD/full-shape analyses, which tend to indicate lower values. A quantitative comparison with the $\Lambda$CDM reference values (Table I in~\cite{Chen:2024vuf}) reveals that our constraints using CMB+DESI-DR2+PPS reduce the $S_8$ tension in full-shape measurements from $4.5\sigma$ to just $1.6\sigma$, representing an approximate 65\% improvement in dataset concordance. It is important to note that the RSD and full-shape analyses are model-dependent, and a re-analysis in this context is necessary for a more meaningful comparison. However, the scenario presented here may still prove to be favorable. A more detailed exploration of these perspectives, considering their complexity, will be presented in future work.

Finally, we can conclude our analysis of the parameters in Table~\ref{tab:Sign-Switching} by examining the transition epoch (represented by $z_{\rm eq, dark}$). The values obtained for this parameter in both data combinations (CMB+DESI-DR2 and CMB+DESI-DR2+PPS) are consistent with constraints from previous studies using DR1 data~\cite{Sabogal:2025mkp}. Moreover, the intriguing negative correlation with $H_0$ and positive correlation with $S_8$ observed in~\cite{Sabogal:2025mkp} also appear in our results, further reinforcing the consistency of the model.

\begin{figure*}[htpb!]
    \centering
    \includegraphics[width=\columnwidth]{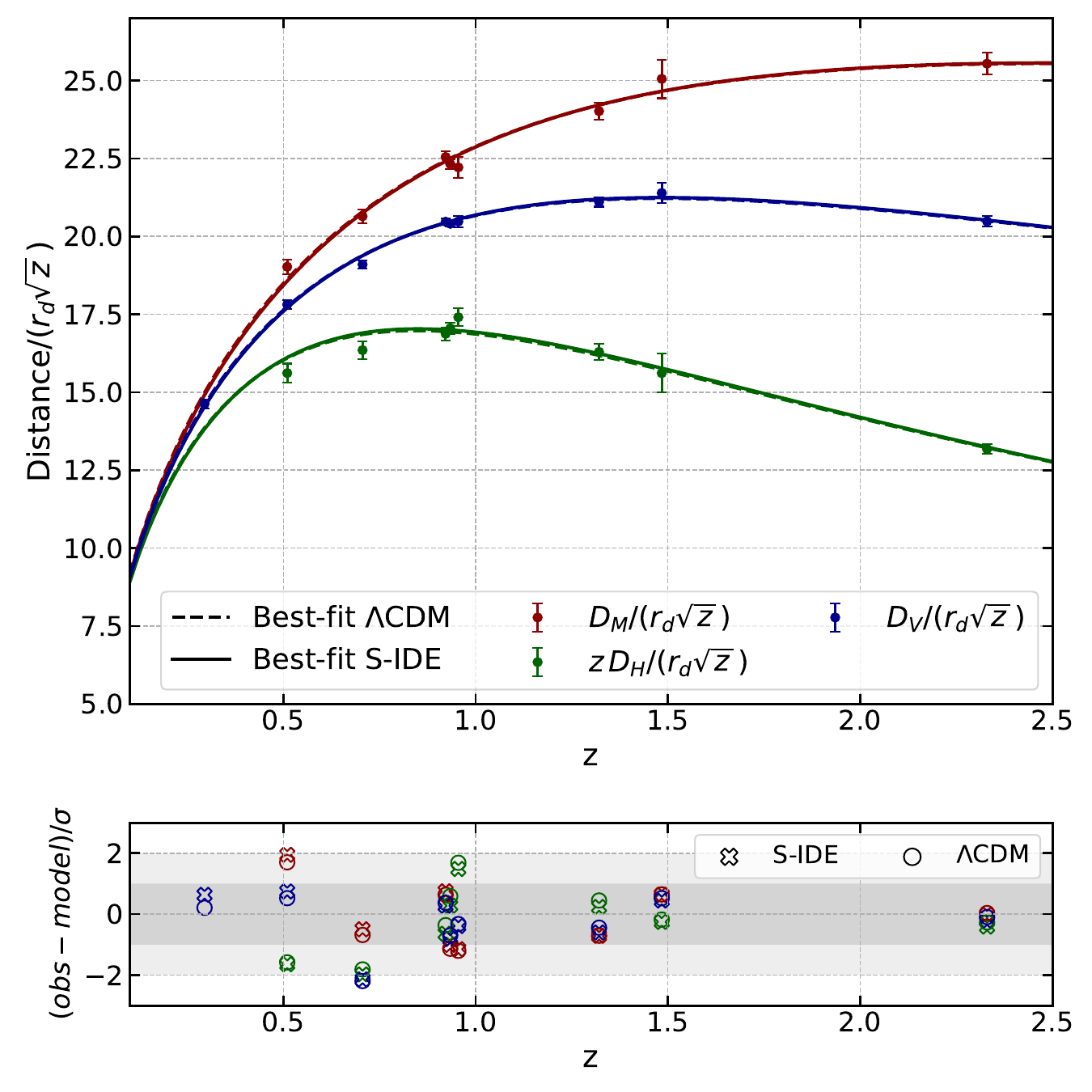} \,\,\,
    \includegraphics[width=\columnwidth]{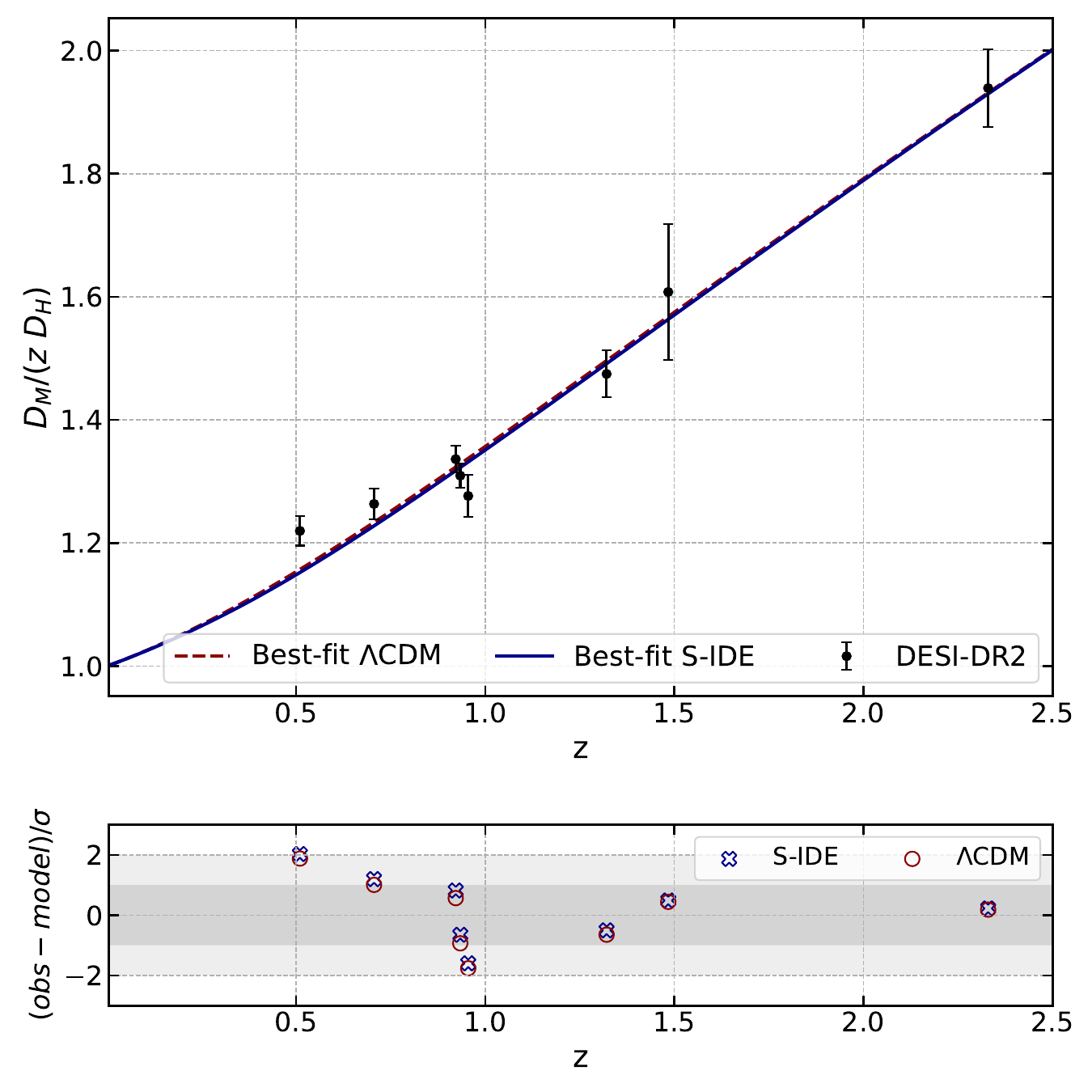}
    \caption{\textit{Upper panel}: Best-fit predictions for the (rescaled) distance–redshift relations for S-IDE (solid curves) and $\Lambda$CDM (dashed curves), obtained from the analysis of Planck-2018+DESI-DR2 data. These predictions are shown for the three types of distances probed by BAO measurements, each indicated by the colors reported in the legend. The error bars represent $\pm 1\sigma$ uncertainties. \textit{Lower panel}: Differences between the model predictions and the data points for each BAO measurement, normalized by the observational uncertainties. The IDE predictions are represented by 'x'-shaped points, while the $\Lambda$CDM predictions are represented by 'o'-shaped points.}
    \label{fig:BAO_1}
\end{figure*}

In order to elucidate the contribution of DESI data, we conducted a comparative analysis between the observed cosmic distances and the theoretical predictions of the S-IDE, IDE, and $\Lambda$CDM models. Figure~\ref{fig:BAO_1} presents: (i) the best-fit predictions derived from the CMB+DESI-DR2 combination for the three rescaled distance variants investigated through BAO measurements; and (ii) the correlation $D_M/(z D_H)$ as a function of redshift for both S-IDE and $\Lambda$CDM models. The lower panel displays the normalized residuals between the DESI observational data and the model predictions—$\Lambda$CDM (represented by circles) and S-IDE (indicated by crosses)—in units of observational uncertainty $\sigma$.
To gain further insight into how S-IDE and $\Lambda$CDM perform when compared against individual BAO measurements, we examine Table~\ref{tab_BAOs}, which displays the theoretical predictions of both models (and traditional IDE) for the four different types of rescaled observables probed by DESI-DR2, evaluated at their respective best-fit parameters obtained from the CMB+DESI-DR2 dataset. For each data point, we show the difference between the predicted and observed values, normalized by the observational uncertainty $\sigma$, thus quantifying the level of agreement or tension.
In general, the S-IDE model exhibits residuals that are comparable to or smaller than those of the $\Lambda$CDM and IDE models in most redshift bins. Notably, at $z = 0.934$ in the $z D_H(z) / (r_d\sqrt{z})$ observable, S-IDE predicts a residual of $0.28\,\sigma$, which is slightly lower than $\Lambda$CDM ($0.59\,\sigma$) and IDE ($0.40\,\sigma$), suggesting a better fit. Similarly, in the $D_M(z)/(r_d \sqrt{z})$ prediction at $z = 0.510$, S-IDE gives a residual of $1.95\,\sigma$, closely matching $\Lambda$CDM ($1.70\,\sigma$) and outperforming the IDE model ($2.46\,\sigma$).
Although all three models show some level of tension with the DESI-DR2 data at specific redshifts, the S-IDE scenario tends to perform consistently well across multiple estimators and redshifts. This suggests that the inclusion of a sign-switching interaction in the dark sector improves the agreement with DESI data, potentially alleviating small-scale discrepancies and complementing global likelihood analyses.
While these differences are modest in isolation, they collectively contribute to the overall improvement in the goodness-of-fit observed in Table~\ref{tab:Sign-Switching}, where the S-IDE model achieves a better fit to the CMB+DESI-DR2 dataset compared to the traditional IDE case, with a reduction in $\Delta \chi^2_{\rm min}$ from $1.1$ to $0.14$, almost statistically indistinguishable from $\Lambda$CDM. This confirms that the overall preference for S-IDE over traditional IDE is partly driven by its enhanced ability to accommodate key features of the new DESI BAO data.

\begin{table}[htpb!]
    \begin{center} \caption{The DESI results (and their $1\sigma$ errors) are presented for four distinct types of distances investigated by BAO measurements. For each data point, we indicate the consistency between the best-fit predictions of the S-IDE, IDE, and $\Lambda$CDM models and the observed data, expressed in units of observational uncertainties ($\#\sigma$).}
        \renewcommand{\arraystretch}{1.25}
        \resizebox{\columnwidth}{!}{
            \begin{tabular}{l c c c c}
                \hline
                \textbf{Distance} & \textbf{Redshift} & \textbf{S-IDE$(\#\sigma)$} & \textbf{$\Lambda$CDM$(\#\sigma)$} & \textbf{IDE$(\#\sigma)$} \\
                \hline
                & & & & \\
                $D_V(z)/(r_d \sqrt{z})$ & $0.295$ & $+0.63 \, \sigma$ & $+0.21 \,\sigma$ & $+1.39\, \sigma$ \\
                & $0.510$ & $+0.74 \, \sigma$ & $+0.52 \,\sigma$ & $+1.44\, \sigma$ \\
                & $0.706$ & $-2.15 \, \sigma$ & $-2.20 \,\sigma$ & $-1.49\, \sigma$ \\
                & $0.922$ & $+0.27 \, \sigma$ & $+0.37 \,\sigma$ & $+0.97\, \sigma$ \\
                & $0.934$ & $-0.81 \, \sigma$ & $-0.70 \,\sigma$ & $-0.02\, \sigma$ \\
                & $0.955$ & $-0.39 \, \sigma$ & $-0.33 \,\sigma$ & $-0.01\, \sigma$ \\
                & $1.321$ & $-0.60 \, \sigma$ & $-0.44 \,\sigma$ & $-0.20\, \sigma$ \\
                & $1.484$ & $+0.45 \, \sigma$ & $+0.52 \,\sigma$ & $+0.62\, \sigma$ \\
                & $2.330$ & $-0.22 \, \sigma$ & $-0.10 \,\sigma$ & $+0.04\, \sigma$ \\
                & & & & \\
                $D_M(z)/(r_d \sqrt{z})$ & $0.510$ & $+1.95 \, \sigma$ & $+1.70 \,\sigma$ & $+2.46\, \sigma$ \\
                & $0.706$ & $-0.50 \, \sigma$ & $-0.68 \,\sigma$ & $+0.04\, \sigma$ \\
                & $0.922$ & $+0.75 \, \sigma$ & $-0.63 \,\sigma$ & $+1.34\, \sigma$ \\
                & $0.934$ & $-1.01 \, \sigma$ & $-1.14 \,\sigma$ & $-0.32\, \sigma$ \\
                & $0.955$ & $-1.15 \, \sigma$ & $+1.20 \,\sigma$ & $-0.84\, \sigma$ \\
                & $1.321$ & $-0.71 \, \sigma$ & $-0.71 \,\sigma$ & $-0.37\, \sigma$ \\
                & $1.484$ & $+0.64 \, \sigma$ & $+0.65 \,\sigma$ & $+0.78\, \sigma$ \\
                & $2.330$ & $-0.02 \, \sigma$ & $+0.04 \,\sigma$ & $+0.20 \,\sigma$ \\
                & & & & \\
                $z D_H(z) / (r_d\sqrt{z})$ & $0.510$ & $-1.64 \, \sigma$ & $-1.58 \,\sigma$ & $-1.49\, \sigma$ \\
                & $0.706$ & $-1.98 \, \sigma$ & $-1.81 \,\sigma$ & $-1.86\, \sigma$ \\
                & $0.922$ & $-0.64 \, \sigma$ & $-0.36 \,\sigma$ & $-0.53 \,\sigma$ \\
                & $0.934$ & $+0.28 \, \sigma$ & $+0.59 \,\sigma$ & $+0.40 \,\sigma$ \\
                & $0.955$ & $+1.48 \, \sigma$ & $+1.68 \,\sigma$ & $+1.56 \,\sigma$ \\
                & $1.321$ & $+0.23 \, \sigma$ & $+0.45 \,\sigma$ & $+0.28 \,\sigma$ \\
                & $1.484$ & $-0.25 \, \sigma$ & $-0.18 \,\sigma$ & $-0.24\, \sigma$ \\
                & $2.330$ & $-0.41 \, \sigma$ & $-0.27 \,\sigma$ & $-0.37\, \sigma$ \\
                & & & & \\
                $D_M(z)/(z D_H)$ & $0.510$ & $+2.03 \, \sigma$ & $+1.88 \,\sigma$ & $+2.15\, \sigma$ \\
                & $0.706$ & $+1.20 \, \sigma$ & $+1.01 \,\sigma$ & $+1.35 \,\sigma$ \\
                & $0.922$ & $+0.81 \, \sigma$ & $+0.57 \,\sigma$ & $+1.02 \,\sigma$ \\
                & $0.934$ & $-0.65 \, \sigma$ & $-0.94 \,\sigma$ & $-0.43 \,\sigma$ \\
                & $0.955$ & $-1.59 \, \sigma$ & $-1.76 \,\sigma$ & $-1.46 \,\sigma$ \\
                & $1.321$ & $-0.50 \, \sigma$ & $-0.65 \,\sigma$ & $-0.36 \,\sigma$ \\
                & $1.484$ & $+0.49 \, \sigma$ & $+0.45 \,\sigma$ & $+0.54 \,\sigma$ \\
                & $2.330$ & $-0.22 \, \sigma$ & $+0.18 \,\sigma$ & $+0.34 \,\sigma$ \\
                \hline\hline
            \end{tabular}
        }
               \label{tab_BAOs}
    \end{center}
\end{table}

\begin{figure}[tpb!]
    \centering
    \includegraphics[width=\columnwidth]{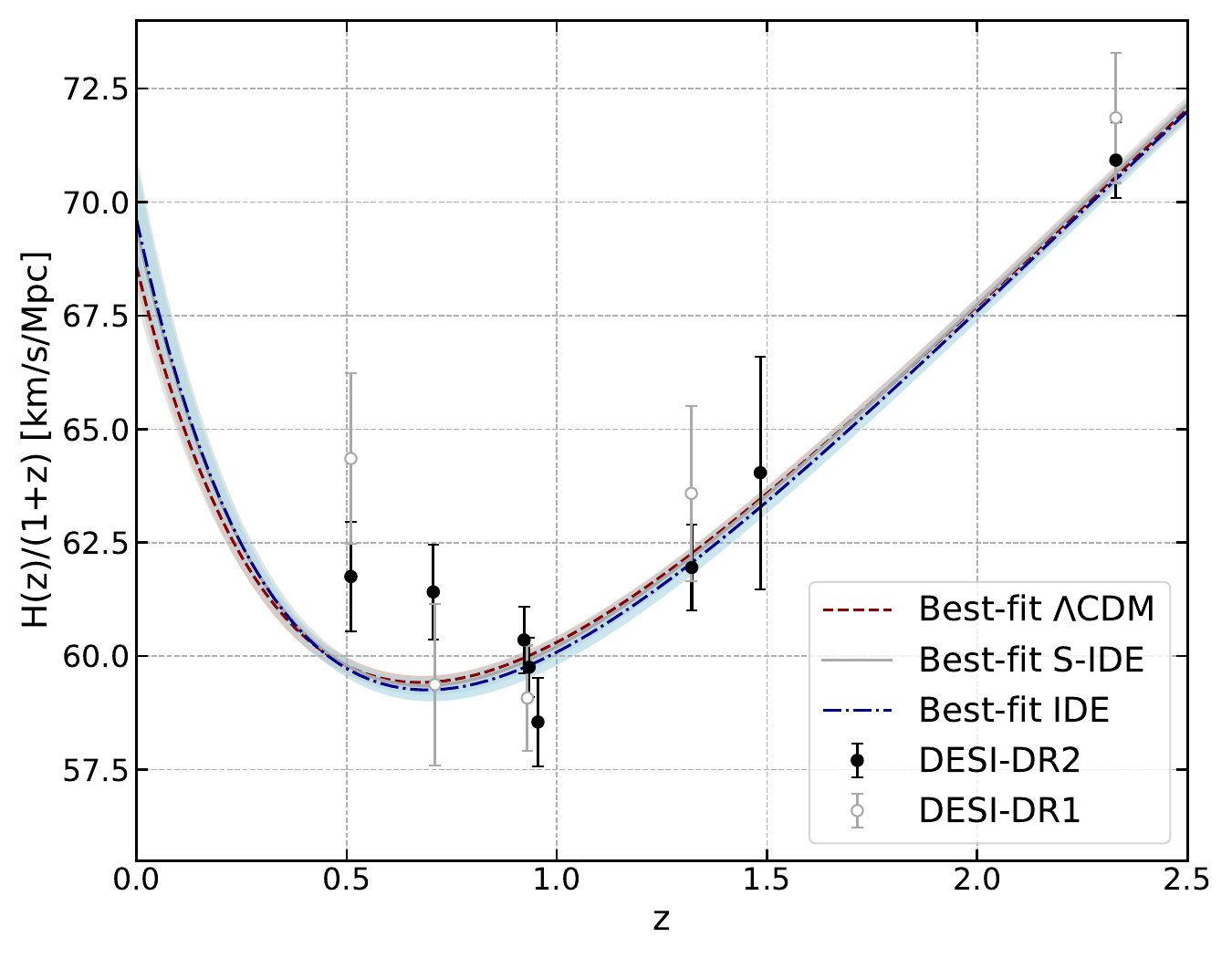} 
    \caption{Statistical reconstruction of the (rescaled) expansion rate of the universe, $H(z)/(1+z)$, at $2\sigma$ confidence levels for the $\Lambda$CDM, S-IDE, and IDE models, based on the joint analysis of CMB+DESI-DR2, compared to DESI-DR2 measurements.}
    \label{fig:H(z)}
\end{figure}

Figure~\ref{fig:H(z)} presents the reconstruction of the $H(z)$ function obtained from the joint CMB+DESI-DR2 analysis. It compares the best-fit curves, along with their $2\sigma$ confidence regions, for the IDE, S-IDE, and $\Lambda$CDM models against the observational DESI data (from both DR1 and DR2). It is evident that the new DR2 measurements exhibit significantly reduced uncertainties compared to DR1, with denser coverage in the low-redshift region ($z < 1$). This improvement allows for a more accurate and precise characterization of the recent cosmic expansion.
Of particular interest are three data points at $z = \{0.510, 0.706, 0.955\}$, which show recurrent discrepancies across all theoretical scenarios considered. These inconsistencies could be indicative of intrinsic limitations within the IDE, S-IDE, and $\Lambda$CDM models, or they may point to unaccounted observational systematics that need further investigation. The observed deviations in these redshift bins highlight the potential challenges in fully capturing the dynamics of cosmic expansion with the current theoretical models and datasets.

The S-IDE scenario exhibits a particularly intriguing dynamical behavior through its effective equation of state:

\begin{equation}
w_{\text{eff}}(z) = -1 + \xi(z) = -1 + \dfrac{\xi_{i}}{3} \operatorname{sgn} \left[ z_{\rm eq,dark} - z \right] \, .
\end{equation}

The observational constraints presented in Table~\ref{tab:Sign-Switching} reveal a well-defined cosmological transition within this model. In the initial phase, corresponding to redshifts greater than $z_{\rm eq,dark}$ ($\xi < 0$), the system resides in the phantom regime, characterized by $w_{\text{eff}} < -1$. As the universe evolves toward lower redshifts ($\xi > 0$), a smooth transition occurs to the quintessence regime, where $w_{\text{eff}} > -1$.

This dynamical feature is in remarkable agreement with recent DESI-DR2 results~\cite{DESI:2025zgx}, where the collaboration analyzed the $(w_0, w_a)$ plane using the parameterization $w(a) = w_0 + w_a(1 - a)$ and found a clear observational preference for the conditions $w_0 > -1$ combined with $w_0 + w_a < -1$. This strongly suggests a universe in which dark energy undergoes an evolutionary transition—exhibiting a phantom-like nature ($w < -1$) in early epochs and adopting quintessence-like characteristics ($w \geq -1$) in more recent cosmic eras, thereby corroborating the physical framework proposed by the S-IDE model.

\section{Final Remarks}
\label{conclusions}

In this work, we have explored the implications of the S-IDE model in comparison with the standard $\Lambda$CDM and traditional IDE frameworks, using the DESI-DR2 measurements of BAO samples. For the first time, for the IDE scenarios considered here, we combine DESI-DR2 data with CMB measurements and the most robust SN samples currently available in the literature. We find that the inclusion of these datasets significantly improves the precision of the cosmological constraints, particularly in the low-redshift region. The data suggest a reduction in the $H_0$ tension to approximately $2.7\sigma$, offering a potential alleviation of the long-standing discrepancy between the local and early-universe measurements of the Hubble constant. However, while the S-IDE model shows promise, it does not provide a complete resolution to the $H_0$ tension, but rather offers significant statistical relief. The behavior of the $S_8$ parameter in the IDE scenario aligns with the latest measurements from the KiDS Legacy, while the S-IDE model permits lower values of $S_8$, which may be consistent with alternative perspectives on the current tension surrounding the $S_8$ parameter from other observational probes.

From a model comparison perspective, we observe that the S-IDE framework exhibits a dynamic evolution in the equation of state, transitioning from a phantom regime to a quintessence-like behavior. This transition aligns with current DESI-DR2 results and presents a novel approach to understanding dark energy and its interactions. However, at low redshifts ($z < 0.7$), the discrepancies between the models (IDE, S-IDE, and $\Lambda$CDM) in explaining the new measurements may indicate intrinsic limitations within the current theoretical frameworks or suggest unaccounted systematic uncertainties in the data.

Looking ahead, future work will focus on extending these analyses by incorporating additional datasets, including the formation of non-linear scale structures, refining model comparisons, and addressing the discrepancies observed at specific redshifts. Improved precision from upcoming observational surveys, such as future DESI releases and complementary probes like \textit{Euclid}, will further enhance our understanding of the potential interaction between dark matter and dark energy.
\\

\textbf{Data Availability}: The datasets and products underlying this research, such as Boltzmann codes and likelihoods, will be available upon reasonable request to the corresponding author after the publication of this article.
\\

\begin{acknowledgments}
The authors thank the referee for the valuable comments and suggestions, which have helped improve the clarity and significance of the results presented in this work. M.A.S and E.S. received support from the CAPES scholarship. E.S. thanks Rogério Riffel for the computational support, which made data production more efficient. M.S. received support from the CNPq scholarship. R.C.N. thanks the financial support from the Conselho Nacional de Desenvolvimento Científico e Tecnologico (CNPq, National Council for Scientific and Technological Development) under the project No. 304306/2022-3, and the Fundação de Amparo à Pesquisa do Estado do RS (FAPERGS, Research Support Foundation of the State of RS) for partial financial support under the project No. 23/2551-0000848-3. E.D.V. is supported by a Royal Society Dorothy Hodgkin Research Fellowship. S.K. gratefully acknowledges the support of Startup Research Grant from Plaksha University  (File No. OOR/PU-SRG/2023-24/08). This article is based upon work from the COST Action CA21136 ``Addressing observational tensions in cosmology with systematics and fundamental physics'' (CosmoVerse), supported by COST (European Cooperation in Science and Technology).

\end{acknowledgments}

\bibliographystyle{apsrev4-1}
\bibliography{main}

\end{document}